\documentclass{iau}

\usepackage{aastex_macros}
\usepackage{graphicx} 
\usepackage{url}

\newcommand{\MHz}{\textrm{MHz}}
\newcommand{\Msun}{\ensuremath{\mathrm{M_{\Sun}}}}
\newlength\abovecaptionskip
\newlength\belowcaptionskip
\title[IAUS291.~~Green Bank Telescope pulsar searches]{The hunt for
  new pulsars with the Green Bank Telescope} \author[R.~S.~Lynch et
  al.]
{Ryan S.\ Lynch$^1$, \\ on behalf of the GBT $\mathbf{350\; \MHz}$
  Drift-scan survey$^2$  \\and Green Bank North Celestial Cap survey$^3$
  collaborations}

\affiliation{$^1$Department of Physics, McGill University, \\ 3600
  University Street, Montreal, QC H3A 2T8, Canada \\ email: {\tt
    rlynch@physics.mcgill.ca} \\ [\affilskip]
  $^2$\url{http://www.as.wvu.edu/~pulsar/GBTdrift350/} \\ [\affilskip]
  $^3$\url{http://arcc.phys.utb.edu/gbncc/}}

\pubyear{2012}
\volume{291}  
\jname{\mbox{Neutron Stars and Pulsars: Challenges and Opportunities
    after 80 years}}
\editors{J. van Leeuwen, ed.} 
\begin{document}

\maketitle

\begin{abstract}
  The Green Bank Telescope (GBT) is the largest fully steerable radio
  telescope in the world and is one of our greatest tools for
  discovering and studying radio pulsars.  Over the last decade, the
  GBT has successfully found over 100 new pulsars through large-area
  surveys.  Here I discuss the two most recent---the GBT $350\; \MHz$
  Drift-scan survey and the Green Bank North Celestial Cap survey.
  The primary science goal of both surveys is to find interesting
  individual pulsars, including young pulsars, rotating radio
  transients, exotic binary systems, and especially bright millisecond
  pulsars (MSPs) suitable for inclusion in Pulsar Timing Arrays, which
  are trying to directly detect gravitational waves.  These two
  surveys have combined to discover 85 pulsars to date, among which
  are 14 MSPs and many unique and fascinating systems.  I present
  highlights from these surveys and discuss future plans.  I also
  discuss recent results from targeted GBT pulsar searches of globular
  clusters and Fermi sources.  
\end{abstract}


\firstsection 
\section{Introduction}
\label{sec:intro}

Since their discovery 45 years ago, just over 2000 pulsars have been
uncovered, enabling some of the most precise and fascinating
astronomical discoveries over that same time span.  The majority of
these are radio pulsars that have been found in large-area surveys,
though a significant fraction, especially of millisecond pulsars
(MSPs), were found in targeted searches.  Almost all surveys result in
some new and often unexpected discovery, many of which have an impact
beyond the study of neutron stars.  Some highlights from only the past
few years are: the double pulsar \cite[(Lyne\,\etal\,2004)]{lbk+04}, which has provided
some of the best tests of strong-field general relativity
\cite[(Kramer\,\etal\,2006)]{ksm+06}, as well pulsar-white dwarf systems that place
stringent limits on tensor-vector-scalar theories of gravity
\cite[(Bhat\,\etal\,2008; Lazaridis\,\etal\,2009; Freire\,\etal\,2012)]{bbv+08,lwj+09,fwe+12}; PSR J1614$-$2230, a $2\; \Msun$ MSP that
has provided the best constraints yet on the equation of state of
ultra-dense matter \cite[(Demorest\,\etal\,2010)]{dpr+10}; the unexpected discovery of a
population of gamma-ray pulsars and MSPs \cite[(Abdo\,\etal\,2009a; Abdo\,\etal\,2009b)]{aaa+09a,aaa+09b}; the
discovery of rotating radio transients (RRATs) and the implication
that they may outnumber traditional radio pulsars \cite[(McLaughlin\,\etal\,2006)]{mll+06}; and
three radio-emitting magnetars \cite[(Camilo\,\etal\,2006; Camilo\,\etal\,2007; Levin\,\etal\,2010)]{crh+06,crhr07,lbb+10} that have
shed light on the relationship between radio pulsars and magnetars.
One of the most exciting prospects for the future is the direct
detection of gravitational waves using a pulsar timing array (PTA) of
ultra-high precision MSPs, which will help to usher in a new era of
gravitational wave astronomy.

New pulsar surveys are important for continuing this tradition of
discovery, especially in light of the need for more high-precision
MSPs for use in PTAs.  Different surveys often complement each other
owing to different sky coverage and observing set-ups (e.g. high
frequency surveys are more sensitive to distant pulsars with high
dispersion measures (DMs), while low-frequency surveys are more
sensitive to steep-spectrum pulsars).  The National Radio Astronomy
Observatory\footnote{NRAO is a facility of the National Science
  Foundation operated under cooperative agreement by Associated
  Universities, Inc.} (NRAO) in Green Bank, West Virginia has a long
history of discovering fascinating pulsars through large-area, low
radio frequency surveys\footnote{I am grateful to Jason Boyles for
  compiling this list.\hfill~}.  The first of these low-frequency Green Bank
pulsar surveys was carried out using the 92-m telescope by
\cite{dth78} at an observing frequency of $400\; \MHz$ and uncovered
17 pulsars.  This was followed up by \cite{dtws85} and
\cite{stwd85}, also using the 92-m telescope at $390\; \MHz$.  The
latter survey was sensitive to pulsars with periods $4\; \mathrm{ms}
\leq P \leq 100\; \mathrm{ms}$ but did not find any, leading the
authors to conclude that the number of such pulsars must be small
compared to long-period pulsars and providing early evidence that the
then-newly discovered MSPs constituted a separate population with a
distinct evolutionary history.  \cite{snt97} discovered a MSP
(defined here as having $P < 20\; \mathrm{ms}$) and a $P = 40.9\;
\mathrm{ms}$ relativistic binary using the 43-m telescope at $370\;
\MHz$.

The Robert C.\ Byrd Green Bank Telescope (GBT) was built to replace
the 92-m telescope and is the largest fully steerable telescope in the
world.  Over the past decade, the GBT has proven to be one of the best
(and perhaps even \emph{the} best) pulsar telescopes in the world,
having discovered over 200 pulsars, many of which are MSPs.  This is
due to its excellent sensitivity and suite of low-noise receivers, its
location in an environment with relatively little radio frequency
interference (RFI), and state-of-the art pulsar back-ends.  The first
low-frequency large-area survey carried out with the GBT was the
$350\; \MHz$ North Galactic Plane survey \cite[(Hessels\,\etal\,2008)]{hrk+08}, which
discovered 33 pulsars.  The GBT has also revolutionized the field of
globular cluster (GC) pulsars, uncovering 71 GC MSPs, more than any
other telescope.  It has also been instrumental in identifying radio
counterparts to Fermi-selected pulsars and MSPs.  Here, I will discuss
the two most recent large-area GBT surveys, the GBT $350\; \MHz$
Drift-scan survey and the Green Bank North Celestial Cap (GBNCC)
survey.  I will also briefly discuss the GBT's role in targeted
searches of GCs and Fermi sources.

\section{The $\mathbf{350\; \MHz}$ Drift-scan Survey}
\label{sec:drift}

\begin{figure}[t]
  \begin{minipage}[b]{0.73\textwidth}
    \centering
    \includegraphics[width=\textwidth]{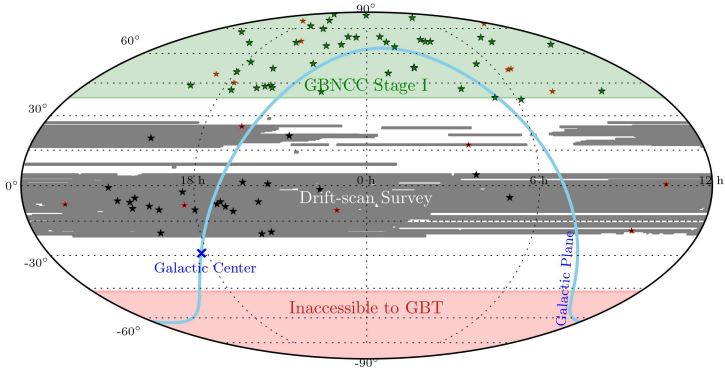}
  \end{minipage}
  \begin{minipage}[b]{0.24\textwidth}
    \caption{Sky coverage of the Drift-scan and GBNCC Stage I surveys.
      New pulsars are shown as stars, and MSPs are outlined in red.
      The GBNCC survey is currently being extended to lower
      declinations with the goal of eventually covering the entire sky
      visible to the GBT.}
  \end{minipage}
  \label{fig:surveys}
\end{figure}

The Drift-scan survey was completed between 2007 May and August while
the GBT was closed for normal operations due to repair of the azimuth
track.  The main science motivation for the survey was the discovery
of high-precision MSPs suitable for inclusion in the North American
Nanohertz Observatory for Gravitational
Waves\footnote{\url{http://nanograv.org}\hfill~} (NANOGrav), as well as other
exotic pulsars.  It covered a variety of right ascensions between
declinations $-7.7^\circ \leq \delta \leq 38.4^\circ$ and $-20.7^\circ
\leq \delta \leq 38.4^\circ$ (see Figure \ref{fig:surveys}) using
$50\; \MHz$ of bandwidth and 140-s long sections of data.  In total,
$10347\; \mathrm{deg^2}$ were observed\footnote{$2800\;
  \mathrm{deg^2}$ were reserved for the Pulsar Search Collaboratory
  \cite[(Rosen\,\etal\,2010)]{rhm+10}.  See
  \url{http://www.pulsarsearchcollaboratory.org/}}, totaling some
$134\; \mathrm{TB}$.  Further details of the survey coverage, data
processing, and sensitivity can be found in \cite{blr+12} and
\cite{lbr+12}.  Data processing for the Drift-scan survey is now
complete, and 35 pulsars have been discovered, including 7 MSPs and
recycled pulsars.  Twenty-four pulsars from early data processing are
presented in \cite{asr+09}, \cite{blr+12}, and \cite{lbr+12} along
with complete timing solutions.  An additional 11 pulsars have been
discovered since this first round of detailed follow-up and are still
being studied.  Below I provide a few highlights.

\begin{itemize}
\item{\textbf{PSR J0348$+$0432} is a 39-ms recycled pulsar in a short,
    2.4-hr orbit around a white dwarf companion \cite[(Lynch\,\etal\,2012)]{lbr+12}.
    Optical imaging and spectroscopic follow-up of the companion have
    provided tight constraints on the mass of the white dwarf
    \cite[(Antoniadis \etal, in prep)]{afw+12}, which is about $0.17\; \Msun$.  The different
    binding energies of the pulsar and white dwarf are predicted by
    certain tensor-vector-scalar gravitational theories to lead to
    strong dipolar gravitational wave emission.  Timing using the
    Arecibo telescope is already placing stringent limits on these
    theories in an as-yet unexplored strong-field regime
    \cite[(Antoniadis \etal, in prep)]{afw+12} and future timing will improve these constraints.}
\item{\textbf{PSR J0337$+$1715} is a 2.7-ms MSP in a hierarchical
    triple, the first to be discovered in the field of the Galaxy.
    The inner companion appears to be a white dwarf with an outer
    companion of undetermined nature on a much longer orbit orbit.
    This system was only recently discovered during final processing
    of the Drift-scan data, but early timing is already showing
    evidence for secular changes to the orbital parameters of the
    inner binary due to three-body interactions \cite[(Ransom \etal, in prep)]{rsh+12}, making
    this system a precision dynamical laboratory.}
\item{\textbf{PSR J2222$-$0137} is a 33-ms recycled pulsar with a
    minimum companion mass of $1.1\; \Msun$ \cite[(Boyles\,\etal\,2012)]{blr+12}.  The
    pulsar is nearby ($D = 310\; \mathrm{pc}$) and bright
    ($S_\mathrm{820\; \MHz} \sim 2\; \mathrm{mJy}$), and has been
    studied using the Very Long Baseline Array \cite[(Deller \etal, in prep)]{d+12}.  The
    pulsar has also been the subject of a campaign to measure Shapiro
    delay \cite[(Boyles \etal, in prep)]{b+12}.  These results will be presented in future
    publications.}
\item{Two \textbf{high-precision MSPs} discovered in the Drift-scan
    survey have been released to NANOGrav and the International Pulsar
    Timing array.}
\item{The well-known ``missing link'' MSP, \textbf{PSR J1023$+$0038}
    \cite[(Archibald\,\etal\,2009)]{asr+09} was one of the earliest Drift-scan discoveries and
    has shed light on the connection between low-mass X-ray binaries
    and MSPs.}
\item{\textbf{Thirty-three RRAT} candidates have been discovered in
    the Drift-scan, with about six having already been confirmed.
    This marks a substantial increase in the size of the RRAT
    population (for details see the discussion by Karako-Argaman, these
    proceedings).}
\end{itemize}

\section{The Green Bank North Celestial Cap Survey}
\label{sec:gasp}

The GBNCC survey is the successor to the Drift-scan survey and was
also carried out at $350\; \MHz$, giving it excellent sensitivity to
nearby, steep spectrum pulsars.  It uses twice the bandwidth of the
Drift-scan survey ($100\; \MHz$), 120-s pointed integrations, and the
newer Green Bank Ultimate Pulsar Processor back-end \cite[(DuPlain\,\etal\,2008)]{drd+08},
which provides increased dynamic range, bandwidth, and better
resistance to RFI (though there is already little RFI).  The science
goals are the same, but with a particular emphasis on northern
declinations, where there are fewer high-precisions MSPs known,
especially in the first stage of the survey.  This is important for
increasing the number of wide-separation baselines in PTAs and also
probes a region of the Galaxy that has not been studied in as much
detail as the Galactic plane.  Stage I of the survey covered the north
celestial cap ($\delta > 38^\circ$) and data taking was completed in
2011.  We are currently conducting the second stage of the survey by
moving towards lower declinations with the goal of eventually covering
the entire sky visible to the GBT.

Data processing is being carried out at the Texas Advanced Computing
Center\footnote{\url{http://www.tacc.utexas.edu/}\hfill~} and especially
using the Guillimin supercomputer operated by
CLUMEQ\footnote{\url{http://www.clumeq.ca/index.php/en/about/computers/guillimin}\hfill~}.
These substantial computing resources (up to 2048 CPUs are reserved
for pulsar searching on Guillimin) have allowed us to reduce most of
the data collected thus far, though we have only managed to do a
preliminary analysis of the resulting pulsar candidates.  These early
results, however, are quite promising, with 50 new pulsars\footnote{Of
  the 50 new pulsars, 21 were identified by high school students
  through a summer internship program at McGill University, and 11
  were identified by undergraduate students as part of the Advanced
  Arecibo Control Center program at the University of Texas at
  Brownsville and the University of Wisconsin-Milwaukee, demonstrating
  how education and outreach programs can be blended with real-world
  research.}, including 9 new MSPs\footnote{See
  \url{http://arcc.phys.utb.edu/gbncc/} for an up-to-date list.\hfill~}.
Detailed follow-up of these new pulsars is ongoing, but I provide some
early highlights below.

\begin{figure}[t]
  \begin{minipage}[b]{0.62\textwidth}
    \centering
    \includegraphics[width=\textwidth]{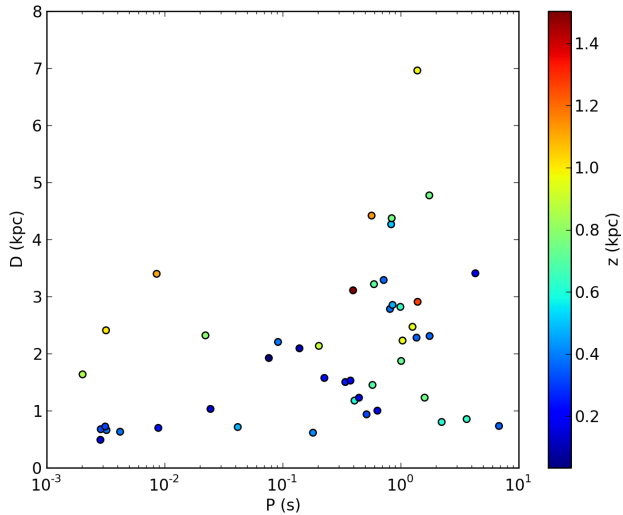}
  \end{minipage}
  \begin{minipage}[b]{0.36\textwidth}
    \caption{Spin periods and DM-inferred distances of newly
      discovered GBNCC pulsars.  The color bar indicates the distance
      above the Galactic plane.  Nearly a dozen MSPs have DM-inferred
      distances $\lesssim$\,1\,kpc, making them promising
      candidates for astrometric and multiwavelength study.}\vspace{2mm}
  \end{minipage}
  \label{fig:gbncc_psrs}
\end{figure}

\begin{itemize}
\item{\textbf{MSPs:} Three of the nine new MSPs appear to be excellent
    candidates for PTAs, and indeed, the 8.85-ms MSP PSR J0645$+$51
    has already been released to NANOGrav and the IPTA.  At least five
    MSPs are in binary systems.  PSR J1816$+$4510 has a $> 0.16\;
    \Msun$ companion and has been detected with Fermi.  There is also
    a UV/optical counterpart \cite[(Kaplan\,\etal\,2012)]{ksr+12}.  PSR J0636$+$51 is a
    2.9-ms MSP in a 1.5-hr orbit.  The lower limit on the companion
    mass is a factor of two higher than that of the so-called
    ``diamond planet'' \cite[(Bailes\,\etal\,2011)]{bbb+11}.  PSR J1124+78 is a nearby ($D
    \sim 600\; \mathrm{pc}$) binary MSP with a likely white dwarf
    companion, and we have proposed for time to look for an optical
    counterpart.}
\item{\textbf{Intermediate Period Pulsars:} Five pulsars have periods
    $20\; \mathrm{ms} \leq P \leq 100\; \mathrm{ms}$, and four have
    Galactic latitudes $|b| > 10^\circ$.  These are properties often
    seen in double-neutron star systems, and there is a possibility
    that at least one of these intermediate period pulsars is in a
    relativistic binary.  The fifth has $b \approx 1^\circ$, and its
    proximity to the plane raises the possibility that it is a young
    pulsar, though no supernova remnants have been found at the
    position of the pulsar in radio and X-ray catalogs.  Further
    investigation is needed to determine the nature of these pulsars.}
\item{\textbf{Nearby Pulsars:} Eleven pulsars (including five MSPs)
    have DM-inferred distances $< 1\; \mathrm{kpc}$ (see Figure
    \ref{fig:gbncc_psrs}) according to the NE2001 free electron
    density model \cite[(Cordes \& Lazio2002)]{cl02}, and several of these have high $350\;
    \MHz$ flux densities.  These are excellent candidates for
    obtaining parallax and proper motion measurements using very-long
    baseline interferometry, as well as for searching for counterparts
    at other wavelengths.  One source in particular, PSR J0737$+$69,
    has a period of $6.8\; \mathrm{s}$, making it the fifth slowest
    rotating radio pulsar known (though there are several magnetars
    and X-ray dim isolated neutron stars with similar or longer
    periods\footnote{See the ATNF pulsar database
      (\url{http://www.atnf.csiro.au/people/pulsar/psrcat/}) and the
      McGill SGR/AXP Online Catalog
      (\url{http://www.physics.mcgill.ca/~pulsar/magnetar/main.html}).}).
    It's proximity makes it an excellent candidate for X-ray
    follow-up.}
\end{itemize}

A more detailed analysis of pulsar candidates from the GBNCC survey,
as well as ongoing data-taking and processing, will undoubtedly result
in the discovery of many more pulsars in the near future.

\section{Targeted Pulsar Searches with the GBT}
\label{sec:targeted}

The GBT has proven itself to be a superb instrument for finding
pulsars through targeted searches, notably of GCs and unidentified
Fermi sources.  Since 2004, nearly all new GC pulsars have been found
with the GBT.  In fact, the GBT has discovered 71 GC pulsars, nearly
half of the entire known population\footnote{See
  \url{http://www.naic.edu/~pfreire/GCpsr.html} for an up-to-date
  list.\hfill~}.  The most promising GCs visible from the GBT have all been
deeply searched \cite[(Ransom\,\etal\,2004; Ransom\,\etal\,2005; Freire\,\etal\,2008; Lynch \& Ransom2011a; Lynch\,\etal\,2011b; Lynch\,\etal\,2012; Stairs \etal, in prep)]{rsb+04,rhs+05,frb+08,lr11,lrfs11,lfrj12,s+12},
with highlights including 34 pulsars in Terzan 5 \cite[(Ransom\,\etal\,2005)]{rhs+05}, the
fastest known rotator \cite[(Hessels\,\etal\,2006)]{hrs+06} and potential super-massive
neutron stars \cite[(Freire\,\etal\,2008)]{fwbh08}.  Though the pace of discovery has slowed
in recent years, GC pulsars are still being discovered with the GBT.
For a more detailed discussion of GC pulsars, see the discussion by
Paulo Freire in these proceedings.

Fermi has revolutionized the search for field radio pulsars,
particularly MSPs.  Radio emission from 45 MSPs and four long-period
pulsars have been discovered by targeting bright, unidentified Fermi
sources, and 26 of these have been found using the GBT (see
\cite{rap+12} for a complete list of Fermi MSPs and their discovery
telescopes).  See the discussions by Pablo Saz Parkinson and Lucas
Guillemot in these proceedings for a more detailed discussion of Fermi
pulsar searches.

\section{Conclusions}
\label{sec:conclusions}

Exciting and unique pulsars are still being discovered almost anywhere
they are looked for, and the GBT continues to play an indispensable
role in this.  The Drift-scan and GBNCC surveys have combined to
discover 85 pulsars, including 16 MSPs or recycled pulsars, in
addition to 33 strong RRAT candidates.  These include some truly
fascinating sources that are enabling cutting-edge physics and
astronomy.  The continuation of GBNCC survey ensures that more
discoveries are yet to come.  Searches of GC are still sensitivity
limited, and searches of unidentified Fermi sources continue with the
promise of yet more MSPs.  As such, the GBT should continue in its
role as one of the best pulsar telescopes in the world.

\end{document}